\documentclass{edp-conf}
\usepackage{graphicx}
\begin{document}

\TitreGlobal{SF2A 2002}

%%-----------------------------
%%      the top matter
%%-----------------------------
\title{Quintessence and the accelerating universe} 
\author{Philippe Brax} \address{Service de Physique Th\'eorique,
CEA-Saclay, F-91191 Gif/Yvette Cedex}
\author{J\'er\^ome Martin}
\address{Institut d'Astrophysique de Paris, 
98bis Blvd Arago, 75014 Paris}
%\author{Author2, B.}\address{...}
%\author{Author3, C.}\address{...}
%
\runningtitle{Quintessence}
\setcounter{page}{237}
% Keep this line, even if the page will be settled afterwards..
\index{J. Martin}
\index{P. Brax}
% Repeat the authors here, this will help to make the final index

\maketitle
\begin{abstract} 
Observations seem to indicate that our universe is presently 
accelerating due to the presence of dark energy. Quintessence 
represents a possible way to model the dark energy. In these 
proceedings, we briefly review its main properties.
\end{abstract}
%
%%-----------------------------
%%      your text
%%-----------------------------
\section{Introduction}
There is now compelling evidence that dark energy is present in our
universe and represents about $70\%$ of the total energy density, i.e.
a value of $\simeq 10^{-47} \mbox{GeV}^4$~(Perlmutter~1998;
Riess~1998). The challenge is therefore to understand the physical
nature of dark energy. Concerning this question, a crucial ingredient
is that, contrary to dark matter, dark energy must possess a negative
pressure as required in order to have an accelerating universe. At
first sight, the most natural candidate is a positive cosmological
constant. A cosmological constant can be viewed as a fluid with a
constant equation of state parameter (the ratio of its pressure to its
energy density) equal to $\omega =-1$. Then, the equation expressing
the conservation of the energy density, $\dot{\rho }+3H(1+\omega
)\rho=0$, implies that the corresponding energy density is
constant. As it is well-known, explaining the dark energy with a
cosmological constant runs into severe problems. Let us briefly
mention only two of them linked to the constancy of the cosmological
constant energy density~(Weinberg~1989). The first problem is that we
need to generate a large cosmological constant in the early universe
(in order to ensure that a phase of inflation took place at this early
epoch) and at the same time we also need a tiny value at the present
time to explain the acceleration of the universe. This seems
incompatible with the fact that the energy density of a $\Lambda
$-term is constant. A second problem is the so-called coincidence
problem. At high redshifts, say just after inflation $z\simeq
10^{28}$, the energy density of the radiation, the dominant fluid at
those redshifts, is of the order $\rho _{_{\rm R}}\simeq
10^{61}\mbox{GeV}^4$, while of course we still have $\rho _{\Lambda
}(z\simeq 10^{28})\simeq 10^{-47}\mbox{GeV}^4$. This means that in
order to have the correct amount of dark energy today, we need to
fine-tune the initial conditions such that $\rho _{_{\rm R}}/\rho
_{\Lambda }\simeq 10^{110}$, something which is a direct consequence
of the fact that $\rho _{\Lambda }$ is a constant. This seems very
unnatural.

\par

The previous considerations lead us to the conclusion that the dark
energy density must be time-dependent. As already mentioned above, the
dark energy pressure must also be negative. As is well-known from
inflationary model-building, a simple way of  obtaining  these
features is to consider a minimally coupled scalar field $Q$ (named
``quintessence''). 
%The energy density of the scalar field is $\rho
%_Q=\dot{Q}^2/2+V(Q)$ while its pressure reads
%$p_Q=\dot{Q}^2/2-V(Q)$. 
The fact that the scalar field evolves with
time causes the equation of state to be redshift-dependent. In
addition, in a regime where the potential energy dominates the kinetic
energy, one obtains a negative pressure. 
%The difference with inflation
%is that the relevant energy scale of the problem is different. 
The question is now to discuss the shape of the potential $V(Q)$ such that
the previous problems can be addressed.

\section{A brief description of the prototypical model of Quintessence}

\begin{figure}[h]
\centering \includegraphics[width=9cm]{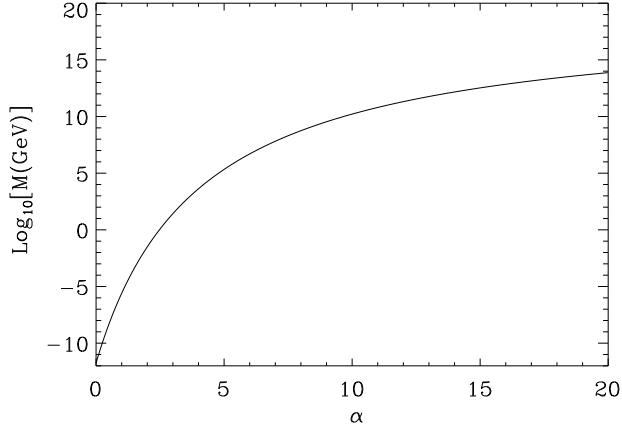} 
\caption{Evolution of the logarithm of the mass scale with the 
power-law index.}  
\label{figure_scalequint}  
\end{figure}

Many potentials have been proposed in the literature. However, a
simple one which allows us to discuss the main features of
quintessence is the so-called Ratra-Peebles potential~(Ratra \& Peebles
1988)
\begin{equation}
V(Q)=\frac{M^{4+\alpha }}{Q^{\alpha }}\, .
\end{equation}
This potential depends on two free parameters: the energy scale $M$
and the index $\alpha $ which is positive. The parameter $\alpha $ is
a priori free whereas the scale $M$ is fixed by the requirement that,
today, $\Omega _Q\simeq 0.7$. One can show that this implies the
following link between $M$ and $\alpha $
\begin{equation}
\log _{10}[M(\mbox{GeV})]\simeq \frac{19\alpha -47}{4\alpha +4}\, .
\end{equation}
This law is plotted in Fig.~\ref{figure_scalequint}. One sees that for
$\alpha >3$, the scale $M$ is beyond the electro-weak scale. In this
case, the inverse power-law shape of the Ratra-Peebles potential
allows us to explain a very small scale in terms of a high-energy scale.

\par

\begin{figure}[h]
   \centering
\includegraphics[width=9cm]{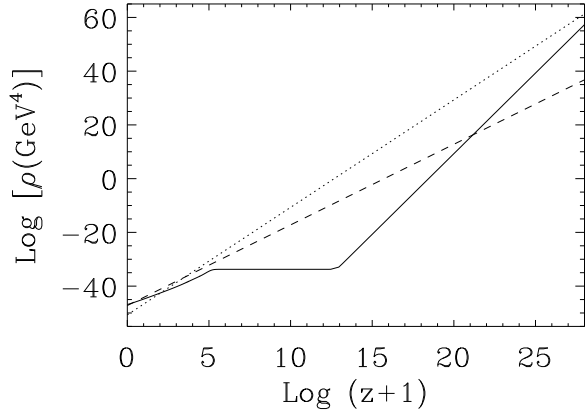}
      \caption{Evolution of the energy density of radiation, matter and 
quintessence with the redshift. The dotted line represents radiation, 
the dashed line, matter, and the solid line, quintessence. The initial 
conditions corresponds to equipartition, i.e. initially 
$\Omega _Q\simeq 10^{-4}\Omega _{_{\rm R}}$.}
       \label{figure_rho_0_17}
   \end{figure}

\begin{figure}[h]
   \centering 
\includegraphics[width=9cm]{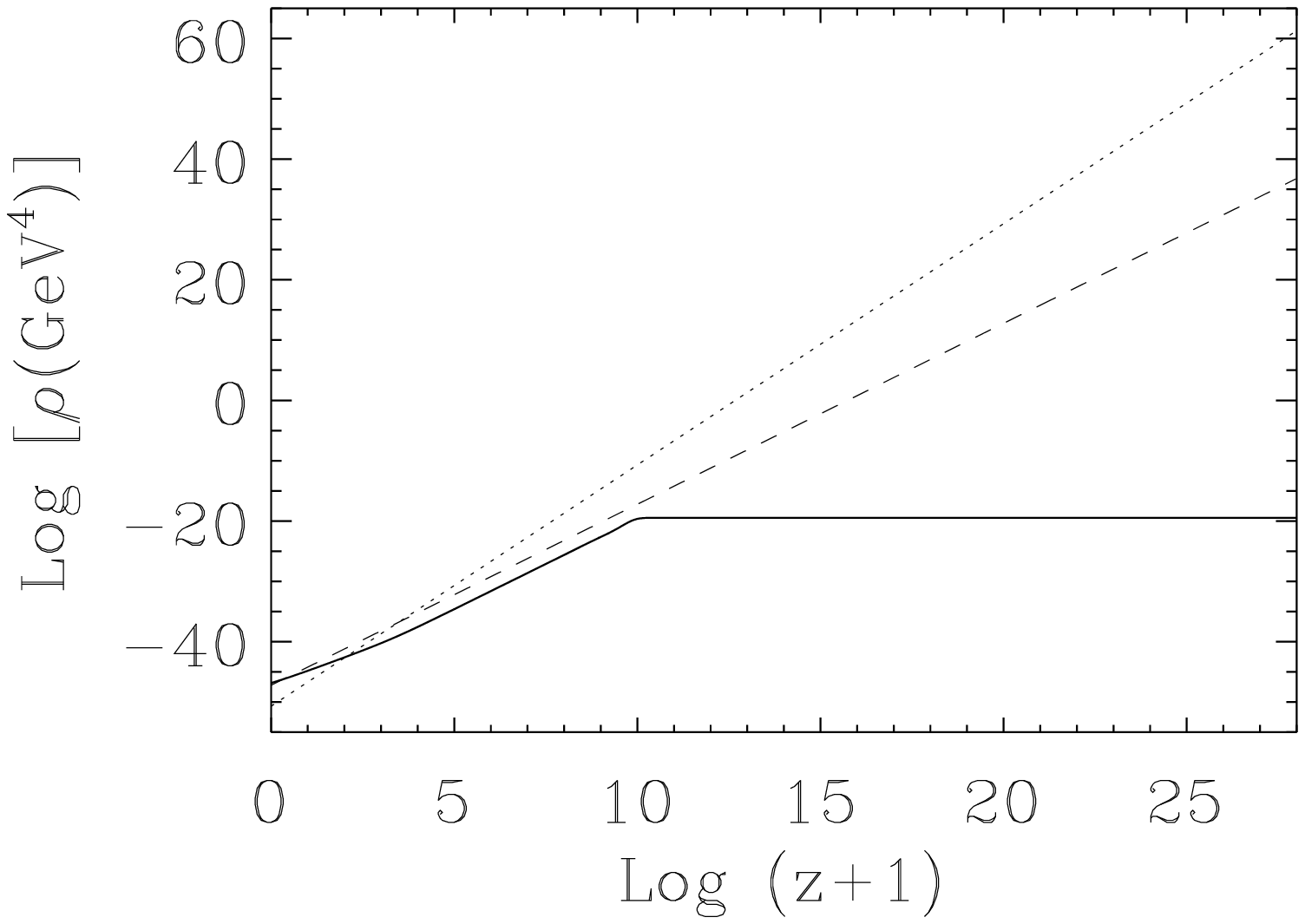} \caption{Same as 
Fig.~\ref{figure_rho_0_17} but with different initial conditions 
such that initially $\rho _Q\ll \rho _{_{\rm R}}$.}  
\label{figure_rho_0_4} 
\end{figure}

We now turn to the coincidence problem evoked in the introduction. In
the case of quintessence, this problem is solved because there exists
an attractor. This means that, whatever the initial conditions, the
solution always converges towards the same solution. Therefore,
contrary to the case of a cosmological constant, there is no need to
fine-tune the initial conditions. This property is illustrated in
Figs.~\ref{figure_rho_0_17} and \ref{figure_rho_0_4}. Starting from
very different initial conditions (different by many orders of
magnitude), one reaches the same solution at small redshifts.
On the attractor, the quintessence energy density redshifts as $ \rho
_Q\propto a^{-3\alpha (1+\omega _{_{\rm B}})/(2+\alpha )}$, where
$\omega _{_{\rm B}}$ is the equation of state of the dominant
hydrodynamical fluid before the quintessence era, i.e. either
radiation ($\omega _{_{\rm B}}=1/3$) or matter ($\omega _{_{\rm
B}}=0$). Since the quintessence energy density redshifts more slowly
than the background, this explains why quintessence starts dominating
at small redshifts while it is hidden during most of the cosmological
evolution.

\par

Finally, let us consider the problem of the equation of state. One can
show that the present value of $\omega _Q$ is such that $-1<\omega
_Q<0$. Due to the presence of the attractor, this value does not
depend on the initial conditions. On the contrary, $\omega _Q$ depends
on the power-law index $\alpha $. For $\alpha =6$, one finds $\omega
_Q\simeq -0.4$ which may be a problem since the observations seems to
indicate that $\omega _Q$ is quite close to $-1$. This can be easily
improved. From a high-energy physics point a view, the SUGRA
model~(Brax \& Martin 1999)
\begin{equation}
V(Q)=\frac{M^{4+\alpha }}{Q^{\alpha }}{\rm e}^{\kappa Q^2/2}\, ,
\end{equation}
with $\kappa =8\pi /m_{_{Pl}}^2$, is better motivated. Due to the
presence of the exponential factor (SUGRA corrections) the equation of
state is pushed towards $-1$, more precisely, one has $\omega _Q\simeq
-0.82$.

\section{Conclusions}

In this short letter, we have quickly reviewed the main properties of
the quintessential scenario for dark energy. This scenario can improve
the fine-tuning problem and solve the coincidence
problem. Quintessence is a falsifiable hypothesis since it predicts
a redshift-dependent equation of state which is different from $-1$
today. Measuring the dark energy equation of state is certainly one of
the most important observational challenge for the future. Only this
measure will able to tell us whether quintessence is an acceptable
explanation of the dark energy of the universe.

%%-----------------------------
%%      your bibliography
%%-----------------------------

\end{document}